\documentclass[11pt,twoside]{article}

\usepackage{baaa2013}

\usepackage{graphicx}
\usepackage{subfigure}
\usepackage{amssymb}
\usepackage[spanish,activeacute,english]{babel}
\usepackage[utf8]{inputenc}
\usepackage{verbatim}
\usepackage{amsmath}
\usepackage{amsfonts}
\usepackage{amssymb}
\usepackage[colorlinks=true,dvips]{hyperref}
\usepackage{natbib}

\begin{document}
\myselectenglish
\vskip 1.0cm
\markboth{Bignone et al.}%
{LGRB Progenitors in Cosmological Simulations}

\pagestyle{myheadings}
\vspace*{0.5cm}


\Categ{2}


\Tema{5}

\vskip 0.3cm
\title{Properties of Long Gamma Ray Burst Progenitors in Cosmological Simulations}

 
\author{L.A. Bignone$^{1}$, L.J. Pellizza$^{2}$ \& P. B. Tissera$^{1}$}

\affil{%
  (1) Instituto de Astronomía y Física del Espacio (CONICET-UBA)\\ 
  (2) Instituto Argentino de Radioastronomía (CONICET)\\ 
}

\begin{abstract} 

  We study the nature of long gamma ray burst (LGRB) progenitors using
  cosmological simulations of structure formation and galactic evolution.
  LGRBs are potentially excellent tracers of stellar evolution in the early
  universe. We developed a Monte Carlo numerical code which generates LGRBs
  coupled to cosmological simulations. The simulations allows us to follow the
  formation of galaxies self-consistently. We model the detectability of LGRBs
  and their host galaxies in order to compare results with observational data
  obtained by high-energy satellites. Our code also includes stochastic
  effects in the observed rate of LGRBs.

\end{abstract}

\begin{resumen}
  
  Estudiamos la naturaleza de los progenitores de estallidos de rayos gamma
  largos (LGRBs) mediante simulaciones cosmológicas de formación de
  estructuras. Los LGRBs son potencialmente excelentes trazadores de la
  evolución estelar en el universo temprano. Desarrollamos un código numérico
  del tipo Monte Carlo que genera LGRBs acoplado a una simulación cosmológica.
  Modelamos la detectabilidad de los LGRBs y sus galaxias anfitrionas
  permitiendo comparar resultados con los datos observacionales obtenidos por
  satélites de alta energía. Nuestro código incluye además efectos
  estocásticos en la tasa de LGRBs observados.

\end{resumen}

\section{Introduction}
\label{sec:introduction}

Long Gamma Ray burst (LGRBS) have long been associated with the last
evolutionary stages of massive stars. This connection coupled with the high
intrinsic luminosity of LGRBs ([$10^{51}$ - $10^{52}$] erg) have prompted
several studies devoted to investigate LGRBS as possible star formation
tracers throughout the universe. However, there are still important questions
regarding the existence of possible biases in the LGRBs population. Observed
LGRBs host galaxies (HGs) appear to be blue and sub-luminous galaxies
\citep{le_floch_are_2003}, and also less massive and poorer in metals than
most star-forming galaxies \citep{savaglio_galaxy_2009}. This suggests the
existence of a more complicated relationship between the production of LGRBs
and the properties of its progenitors. Some authors propose that a chemical
dependence hypothesis can explain the observations
\citep{daigne_redshift_2006, salvaterra_gamma-ray_2007, li_dust_2008}.

One proposed way to study this problem is to compute a simulated LGRB
population (i.e., redshift distribution, peak luminosities, intrinsic spectral
parameters), and compare the predictions of the model to gamma-ray observables
such as the distributions of peak fluxes, redshifts and observed spectral
parameters \citep{daigne_redshift_2006, salvaterra_gamma-ray_2007}. A comoving
LGRB rate proportional to the comoving star formation rate (SFR) is usually
assumed, together with a redshift or metallicity-dependent proportionality
factor.

An alternative approach consists in computing the simulated LGRBs population
within a cosmological simulation of galaxy formation. The great advantage of
this method is that it allows for a consistent treatment of the evolution of
the SFR and the metallicity of stars and also allows for the joint study of
LGRBs progenitors and their host galaxies \citep{nuza_host_2007,
chisari_host_2010, artale_chemical_2011}. In this work we propose a Monte
Carlo code based on the latter method that improves upon previous versions by
taking into account stochastic effects in the observed LGRBs rate due to the
sparse nature of very massive stars. Stochastic effects are a crucial element
for understanding small stellar populations
\citep{da_silva_slugstochastically_2012}.

\section{Numerical procedure}
\label{sec:procedure}

\subsection{Cosmological simulation}
\label{sec:cosmological_simulation}

We use a hydrodynamic cosmological simulation obtained by using a version of
GADGET-3. The simulation is consistent with the concordance $\Lambda$-CDM
model with cosmological parameters: $\Omega_\lambda$ = 0.7, $\Omega_m$ = 0.3,
$\Omega_b$ = 0.04, $\sigma_8$ = 0.9 and H$_0$ = 100 h km s$^{-1}$ Mpc$^{-1}$
with h = 0.7; where $\Omega_\lambda$, $\Omega_m$ and $\Omega_b$ are the
density parameters for dark energy, matter and baryons, respectively;
$\sigma_8$ is the normalization of the matter power spectrum on scales of 8
h$^{-1}$ Mpc and H$_0$ is the Hubble constant.

Initially, the cosmological simulation consists of $230^3$ dark matter
particles with a mass of $\sim8.47\times10^6$ M$_\odot$ and $230^3$ gas
particles with a mass of $\sim1.3\times10^6$ M$_\odot$. Gas particles have an
initial hydrogen and helium proportions given by  XH = 0.76 and XHe = 0.24
respectively, the code then follows the chemical enrichment of: $^1$H, $^2$He, $^{12}$C,
$^{16}$O, $^{24}$Mg, $^{28}$Si, $^{56}$Fe, $^{14}$N, $^{20}$Ne, $^{32}$S, $^{40}$Ca and $^{62}$Zn.

Stellar formation occurs when the inter stellar medium (ISM) density is above
a critical density $\rho_c > 0.032$ g cm$^{-3}$. A multiphase treatment for
gas particles and Supernovae feedback (SNII, SNIa)
\citep{scannapieco_feedback_2006} is also included. The energy feedback to the
ISM per Supernova (in units of 10$^{51}$ erg) is 0.7. The mass of metals that
goes into the cold phase of the ISM in a Supernova explosion is 50\%.

\subsection{Synthetic LGRBs population}
\label{sec:synthetic_lgrbs_population}

In order to simulate the LGRBs population we adopt first a minimum mass
(M$_{\text{min}}$) and a maximum metallicity (Z$_{\text{max}}$) for the LGRBs
progenitors. M$_{\text{min}}$ allows us to restrict the simulated progenitors
to massive stars, while Z$_{\text{max}}$ introduces a metallicity threshold
for the LGRBs progenitors metallicity. We also adopt an initial mass function
(IMF) for the stellar population \citep{chabrier_galactic_2003}.

To take into account the stochastic effects in the rate of LGRBs due to the
small sample resulting in considering only massive stars as progenitors we
developed a numerical code that uses the IMF as a probability distribution in
order to build the stellar population of a stellar particle piece by piece
with the correct stochastic properties. We accomplish this by randomly drawing
masses from the IMF until the total mass of the stellar particle is reached.

For each stellar population formed in the simulated galaxies, represented by a
particle i, we are able to establish the number $N_i$ of LGRBs which are going
to be  produced. Stellar populations that have been enriched beyond
Z$_{\text{max}}$ are discarded.

Once we have established the number of progenitors we are able to assign each
LGRB intrinsic properties, like the isotropic luminosity and the spectrum
parameters. The time-averaged gamma-ray burst spectra can be well described by
the Band function \citep{band_batse_1993} for which we have adopted fixed
values $\alpha = -1$ and $\beta = -2.25$. Each LGRBs spectrum is then
determined solely by the peak energy ($E_p$). The $E_p^j$ value for each LGRB
event ($j$) is assigned from a log-normal distribution with mean value and
standard deviation which are considered free parameters of our model. For the
isotropic luminosity distribution we initially considered a simple power law
with exponent $\nu$ and a luminosity range [$L_\text{min}$ - $L_\text{max}$],
where $\nu$, $L_\text{min}$ and $L_\text{max}$ are also free parameters of our
model.

\section{Results}
\label{sec:results}

The intrinsic energetic properties of the LGRB together with the redshift ($z$) obtained
from the cosmological simulation allows us to calculate the photon peak flux
($P^j$) observed for each event by a high-energy satellite in a given spectral
window [$E_1$ - $E_2$].
\begin{equation}
P^j(L^j, E_p^j) = \frac{(1+z)^{\alpha+2}A\int_{E_1}^{E_2}[B(E)/A]dE}{4\pi d_L^2(z)}, 
\end{equation}
where $B(E)$ is the Band function, $d_L(z)$ is the luminosity distance and $A$ is a normalization factor given by 
\begin{equation}
A = L^j\left(\int_{1\,\text{keV}}^{10^4\,\text{keV}} E[B(E)/A]dE\right)^{-1}.
\end{equation}
This allows us to contrast our results with those obtained by BATSE and SWIFT satellites, which have different spectral observational windows, 50 - 300 keV and 15 - 150 keV respectively.

In the case of BATSE, we have adopted the catalog proposed by \cite{stern_off-line_2001} that establishes a well determined efficiency given by
\begin{equation}
\epsilon_B(P^j) \propto [1 - \exp(-(P/P_\star)^2)]^{\nu_B},
\label{eq:eff}
\end{equation}
where $P_\star = 0.129$ ph s$^{-1}$ cm$^{-2}$ and $\nu_B = 2.34$. This allows us to model the response of the detector by a simple Monte Carlo experiment that accepts events with a probability given by equation \ref{eq:eff}.

In the case of Swift we have compiled data obtained from the public Swift
database up to GRB 130907A. For the efficiency we assume simple a cut at P = 0.4 ph
s$^{-1}$ cm$^{-2}$ because of the difficulty in modeling SWIFT's response
\citep{band_postlaunch_2006}.

Figure 1 compares the peak flux distribution and spectral peak energy
distribution observed by BATSE and SWIFT with results of our simulation for
different values of the metallicity cut $Z_\text{max}$. A good agreement could
be found for $Z_\text{max} = 0.006$ in the case of BATSE, but SWIFT
observations are not well reproduced by any model. These disagreement suggests
that a more complex dependency between LGRBs and the properties of their
progenitors might exist. A dependency between the metallicity and the
luminosity will be explored in the future.

\begin{figure}[!t]
  \centering
  \includegraphics[width=0.9\textwidth]{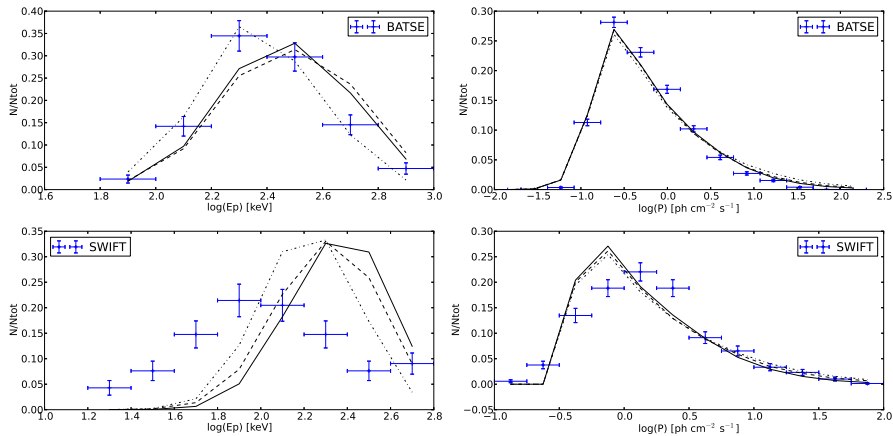}
  \caption{Top left: peak energy distribution observed by BATSE. Bottom left: peak energy distribution observed by SWIFT. Top right: peak flux distribution observed by BATSE. Bottom left: peak flux distribution observed by SWIFT. The lines represent results for our simulation with values of $Z_\text{max}$: 1 (solid), 0.01 (dashed) and  0.006 (dotted)}
  \label{fig:results}
\end{figure}

\bibliographystyle{baaa}
\bibliography{Bignone_BAAA56}

\end{document}